\documentstyle[aps,epsf,twocolumn]{revtex}

\begin{document}
\title{Electron Spin Resonance Transistors for Quantum Computing in \\
Silicon-Germanium Hetero-structures}
\author{Rutger Vrijen$^1$, Eli Yablonovitch$^1$, Kang Wang$^1$, Hong Wen
Jiang$^2$, Alex Balandin$^1$, Vwani Roychowdhury$^1$, Tal Mor$^1$ and David DiVincenzo$^3$ }
\address{
$^1$~University of California, Los Angeles, Electrical Engineering Dept., 
Los Angeles, California\\
$^2$~University of California, Los Angeles, Physics Dept., Los Angeles, California\\
$^3$~IBM T.~J.~Watson Research Center, Yorktown Heights, New York\\
}
\maketitle
\newcommand{\degr}{$^{\circ}$}

\abstract{ 
We apply the full power of modern electronic band structure 
engineering and epitaxial hetero-structures to design a transistor that can sense 
and control a single donor electron spin.  Spin resonance transistors may form the 
technological basis for quantum information processing.  One and two qubit 
operations are performed by applying a gate bias.  The bias electric field pulls 
the electron wave function away from the dopant ion into layers of different 
alloy composition.  Owing to the variation of the $g$-factor (Si:$g$=1.998, 
Ge:$g$=1.563), this displacement changes the spin Zeeman energy, allowing single-
qubit operations.  By displacing the electron even further, the overlap with 
neighboring qubits is affected, which allows two-qubit operations.  Certain 
Silicon-Germanium alloys allow a qubit spacing as large as 200 nm, which is well 
within the capabilities of current lithographic techniques.  We discuss 
manufacturing limitations and issues regarding scaling up to a large size 
computer.
}

\section{Introduction}
\label{sec:introduction}
The development of efficient quantum algorithms for classically hard problems 
has generated interest in the construction of a quantum computer.  A quantum 
computer uses superpositions of all possible input states. By exploiting this 
quantum parallelism, certain algorithms allow one to 
factorize\cite{shor-94} large integers with astounding speed, and 
rapidly search through 
large databases\cite{grover-97}, and efficiently simulate quantum 
systems\cite{feynman-82}.  In the nearer term such devices could facilitate 
secure communication and distributed computing. 

In any physical system, bit errors will occur during the computation.  In 
quantum computing this is particularly catastrophic, because the errors cause 
decoherence\cite{unruh-95} and can destroy the delicate superposition that needs 
to be preserved throughout the computation. With the discovery of quantum error 
correction\cite{shor-95} and fault-tolerant computing, in which these errors are 
continuously corrected without destroying the quantum information, the 
construction of a real computer has become a distinct possibility. 

Even with the use of fault-tolerant computing a quantum computer engineer would 
still prefer a system that exhibits the smallest possible error rate on the 
qubits, the two level systems that hold the quantum information. In fact, 
Preskill\cite{preskill-98} (in a review of the subject) presented a requirement for 
fault-tolerance; the 
ratio of the error rate to the computer clock rate has to be below a certain 
threshold.

Several systems have recently been proposed to obtain a physical implementation 
of a quantum computer.  These systems include cold ion traps\cite{cirac-95}, nuclear 
magnetic resonance (NMR) systems\cite{gershenfeld-97,cory-98}, all-optical logic 
gates\cite{turchette-95,milburn-89}, Josephson junctions\cite{shnirman-97}, and 
semiconductor nanostructures\cite{barenco-95}. Successful experimental 
demonstrations of one and two qubit computers were reported for trapped 
ion systems\cite{monroe-95} and NMR systems\cite{jones-98}. 

Last year, Bruce Kane\cite{kane-98} proposed a very interesting and elegant 
design for a spin resonance transistor (SRT).  He proposed to use the nuclear 
spins of $^{31}$P dopant atoms, embedded in a Silicon host, as the qubits.  At low 
temperatures the dopant atoms do not ionize, and the donor electron remains 
bound to the $^{31}$P nucleus. The control over the qubits is established by 
placing a gate-electrode, the so-called A-gate, over each qubit. By biasing the 
A-gate, one can control the overlap of the bound electron with the nucleus and 
thus the hyperfine interaction between nuclear spin and electron spin, which 
allows controlled one-qubit rotations.  A second attractive 
gate, a J-gate, decreases the potential barrier between neighboring qubits, and 
allows two nuclear spins to interact by electron spin-exchange, which 
provides the required controlled qubit-qubit interaction.

The rate of loss of phase coherence between qubits in a quantum system is 
typically characterized by the dephasing time T$_2$.  The T$_2$ dephasing time 
of the nuclear spins in silicon is extremely long.  The silicon host 
efficiently isolates the nuclear spins from disturbances\cite{divincenzo-96}.
A quantum computer based on semiconductors offers an 
attractive alternative to other physical implementations due to 
compactness, robustness, the potentially large number of 
qubits\cite{bandyopadhyay-98}, and semiconductor compatibility with industrial 
scale processing.  However, the required transistors are very small, since 
their size is related to the 
size of the Bohr radius of the dopant electron. Furthermore, after the calculation 
is completed Kane's SRT requires a sophisticated spin transfer between nuclei and 
electrons to measure the final quantum state.

We suggest using the full power of modern 
electronic band structure engineering and epitaxial growth techniques, to introduce 
a new, more practical, field 
effect SRT transistor design that might lend itself to a near term demonstration of 
qubits on a Silicon wafer.  We alter Kane's approach by the implementation of 
these spin-resonance transistors in engineered Germanium/Silicon hetero-structures 
that have a controlled band structure.  Si-Ge strained 
hetero-structures, developed by IBM and other companies, are in the mainstream of 
Silicon technology, and are currently used for high frequency wireless 
communication transistors, and high-speed applications.

In Si-Ge hetero-structure layers we can control the effective mass of the donor 
electron to reduce the 
required lithographic precision, and to permit the SRT transistors to be as 
large as $\approx 2000$~\AA. The Bohr radius of a bound electron in Si-Ge 
can be much larger than in Silicon due to the very small effective mass in 
strained Si-Ge alloys, and their higher dielectric constant.  This places the 
lithographic burden well within the practical range of electron beam 
lithography and almost within range of contemporary optical lithography.  

Among the other simplifications, we will employ an electron spin, rather than a 
nuclear spin as the qubit. Owing to the difference in the electronic $g$-factor, 
$g=1.998$ for Si, and $g=1.563$ for Ge, the electron spin resonance transition 
can be readily tuned by an electrostatic gate on a compositionally modulated Si-Ge 
epilayer structure.  By working with electron spins rather than nuclear 
spins, we avoid the requirement of a sophisticated spin transfer between 
electrons and nuclei, for read-in/read-out of quantum data and for the operation of 
two-qubit gates.  In addition, due to 
their higher Zeeman energy, electron spins will eventually permit a clock speed 
up to 1~GHz compared to a speed$\approx 75$~kHz projected for the nuclear spins.  
Likewise, isotopic purity is not critical for electron spins.

In order to read-out the final result of a quantum calculation we will need to 
be able to detect single electron charges.  Individual electro-static charges 
are readily detected by conventional field effect transistors (FET's) at low 
temperatures, which obviates the need for the sophisticated single electron 
transistors (SET's). In this paper, we illustrate our design for an electron spin 
resonance transistor.

\section{Electron spin dephasing time in silicon and germanium}
\label{sec:dephasing}
Electron spins benefit from the same protective environment provided by the 
silicon host as nuclear spins.  Indeed, the ESR line in doped Silicon at 
low temperatures turns out to be exceptionally clean and narrow compared to 
other ESR lines.  

Feher\cite{feher-59a,feher-59b,wilson-61} found that the Si:$^{31}$P ESR line is 
inhomogeneously broadened by hyperfine interactions with neighboring nuclear 
spins.  But the nuclear spin flip T$_1$ relaxation times were 
measured\cite{feher-59b} to be in the 1-10~hour range. Thus the nuclei can 
be regarded as effectively static on the time scales needed for quantum 
computing.  Likewise the direct electron spin-flip T$_1$ is also 
around\cite{feher-59b} an hour.  

On the question of the critical transverse T$_2$ ESR dephasing linewidth there 
was only a little information.  Feher and Gere studied some heavily doped n-Si:P 
samples, and found that the ESR linewidth actually narrowed\cite{feher-63} at 
high doping, down to a 1~MHz linewidth at the 9~GHz ESR frequency, for the heavy 
doping level, $n=3 \times 10^{18}$/cm$^3$.  This unusual behavior was clearly the 
result of exchange narrowing of the hyperfine inhomogeneity.  For quantum 
computing, the issue is the linewidth of a single electron spin transition, rather 
than a heavily doped inhomogeneous ensemble. 

Thus the outlook was optimistic.  If the linewidth is only 1~MHz at such a high 
doping level, and is due to exchange with neighboring electrons, then the 
linewidth would surely be much narrower at lower doping levels, and especially 
for one isolated electron.  Indeed that was confirmed by Chiba and 
Hirai\cite{chiba-72} who measured a 1/2$\pi$T$_2$~linewidth of only $\approx$1~kHz at a 
doping of 10$^{16}$~Phosphorus ions per cm$^3$, by the very reliable spin-echo 
technique.  The residual linewidth was interpreted as being due to spin 
diffusion via the nuclear spins.  Indeed the linewidth was shown\cite{gordon-58} 
to narrow further in isotopically purified, 0~spin, Si$^{28}$, making the 
T$_2$~dephasing even slower. The observed 1~kHz linewidth at n=10$^{16}$/cm$^3$ is 
already narrow enough, in relation to the 9~GHz ESR frequency to allow enough
operations for fault tolerant computing\cite{preskill-98}. 

In germanium the dominant mechanism for spin dephasing is quite different from the 
one in silicon. Theory\cite{roth-60,hasegawa-60} and experiment\cite{wilson-64} 
have confirmed that the dominant relaxation in germanium is through acoustic disturbances
of the spin-orbit coupling. The $g$-factor in germanium is much different from 2, the free 
electron 
value, because of the relatively strong spin-orbit coupling. Germanium has four ellipsoidal 
conduction band minima, which are aligned with the $\langle 111 \rangle$ directions. 
In each minimum, the effective mass depends on the direction of electron motion, with a low effective 
mass ($m_{xy}$) in the transverse direction and a high effective mass in the longitudinal
direction ($m_z$)(see Table~\ref{table:table1}). The anisotropic effective mass results in an anisotropic $g$-factor, with 
$g=g_{\parallel}$ for magnetic field components in the $\langle 111 \rangle$ 
direction, and $g=g_{\perp}$ for magnetic field components perpendicular to this direction. 
For arbitrary angles $\phi$ between the magnetic field and the $\langle 111 \rangle$ 
direction the $g$-factor is given by
\begin{equation}
g^2=g_{\parallel}^2 \cos^2 \phi + g_{\perp}^2 \sin^2 \phi
\label{eq:g_phi}
\end{equation}

The electronic ground state of the donor atom is an equal superposition (singlet) state of
the four equivalent conduction band minima, and therefore has an isotropic $g$-factor, 
$g=g_{\parallel}/3+2g_{\perp}/3=1.563$. 
However, in the presence of lattice strain, the energies of the conduction band minima shift 
with respect to each other. In the new donor ground state, probability is shifted among the four
valleys, with some valleys more populated than others. This produces a shift $\Delta g$ in the 
$g$-factor, since each valley forms a different angle $\phi$ with the static magnetic field $B$. 
The corresponding relative energy shift of the spin states is
proportional to $(\Delta g) \mu B$ with $\mu$ the Bohr magneton. At finite temperatures, 
acoustic phonons cause time-varying strains with a finite power density at the spin 
transition energy, which induce spin-lattice relaxation. 

At these temperatures it follows from this theory that the phase relaxation time is
of the same magnitude as the population relaxation time $T_2 \approx T_1$. Experiments have shown
that $T_1$ is around $10^{-3}$ seconds for germanium at 1.2~K. We are not aware of direct 
measurements of $T_2$ by electron spin resonance experiments similar to those that were done in 
silicon. Unless there are other, as of yet unknown $T_2$ 
mechanisms in germanium, the $T_2$ will be determined by acoustic vibrations and 
be of the order of $10^{-3}$ seconds, which is equal to the best measured $T_2$ in silicon, and is 
again sufficiently long to allow fault tolerant computing. 

Several mechanisms could lead to a further improvement in the $T_1$ and $T_2$ caused by acoustic 
vibrations. Firstly, working at lower temperatures will reduce 
the phonon energy density, which is proportional to $T^4$. 
Secondly, for the two orientations of germanium that we propose to use, 
$\langle 111 \rangle$ and $\langle 001 \rangle$, some special considerations can make the expected
lifetimes longer.
For germanium grown with strain in the $\langle 111 \rangle$ direction, the conduction band
minimum along the growth direction has a significantly lower energy than the other three
minima. In the electronic ground state, virtually all population resides in this minimum, and 
there is little coupling to the three split-off valleys. In the theory by Roth and 
Hasegawa\cite{roth-60,hasegawa-60}, this effect is accounted for by a square 
dependence of $T_1$ on the energy splitting between the electronic ground state and excited states 
(singlet-triplet splitting). The grown-in strain increases this splitting from 2 meV to 200 meV, with
a corresponding increase in lifetime of $10^4$.
For germanium grown with strain in the $\langle 001 \rangle$ direction and with the magnetic field aligned
with that direction, a symmetry argument forbids a strain induced $g$-shift: the 
$\langle 001 \rangle$ direction makes equal angles with all conduction band minima, and therefore
a probability redistribution among these minima does not affect the $g$-factor, as can be seen from
Equation~\ref{eq:g_phi}. Thus, further improvements in the already acceptable lifetimes appear
possible.

The electron spin resonance (ESR) of a 
bound donor in a semiconductor host provides many advantages: Firstly, in a magnetic field of 
2~Tesla, the ESR resonance frequency is$\approx$56~GHz, easily allowing qubit 
operations at up to $\approx$1~GHz.  This is comparable to the clock speed of 
ordinary computers, and is consistent with the precision of electronic control 
signals that are likely to be available.  Secondly, at temperatures well below 1~K, 
the electron spins are fully polarized allowing a reproducible starting point 
for the computation.  And finally, for electron spins isotopic purity is not 
compulsory since the nuclear spin inhomogeneity remains frozen at low 
temperatures.

\section{SRT Transistor Size and Lithographic Critical Dimension}
\label{sec:size}
The Bohr radius of the bound carrier wave function regulates the size scale of 
Spin Resonance Transistors.  In semiconductors the Bohr radius is much larger 
than in vacuum, since the Coulomb force is screened by the dielectric constant, 
and the effective mass is much smaller.  Thus the bound carrier roams farther.  
The Bohr radius is:  $a_B=\epsilon \frac{m_0}{m^{\ast}}(\frac{\hbar^2}{m_0 
q^2})$ in the semiconductor, where $\frac{m^{\ast}}{m_0}$ is the effective mass 
relative to the free electron mass, $\epsilon$ is the dielectric constant, 
$\epsilon=16$ for Ge, and $\epsilon=12$ for Si and the quantity in parenthesis 
is the Bohr radius in vacuum.  

It is common in Si-Ge alloys to have strain available as an engineering 
parameter.  Strain engineering of valence band masses has been very successful, 
and is used\cite{yablonovitch-88} in virtually all modern semiconductor lasers.  
As discussed above, in the conduction band, strain splits the multiple conduction band valley 
energies, allowing one valley to become the dominant lowest energy conduction 
band.  If that valley also happens to be correctly aligned, the donor wave 
functions can have a low mass moving in the plane of the silicon wafer, and a 
high mass perpendicular to the wafer surface.  That is exactly what we are 
looking for in spin resonance transistors.  We want large wave functions in the 
directions parallel to the wafer surface, in order to relax the lithographic 
precision that would have been demanded if the Bohr radius were small.

In Si-rich alloys there are 6~conduction band minima, in the 6~cubic directions, 
that are frequently labeled as the X-directions.  In Ge-rich alloys, there are 
4~conduction minima located at the $\langle 111 \rangle$ faces of the Brillouin 
Zone, labeled~L.  The Ge-rich case is particularly interesting, since it has a 
conduction band mass of only $0.082m_0$ in the transverse direction.
\begin{figure}[p]
\centerline{\epsfxsize=250pt
		\epsfbox{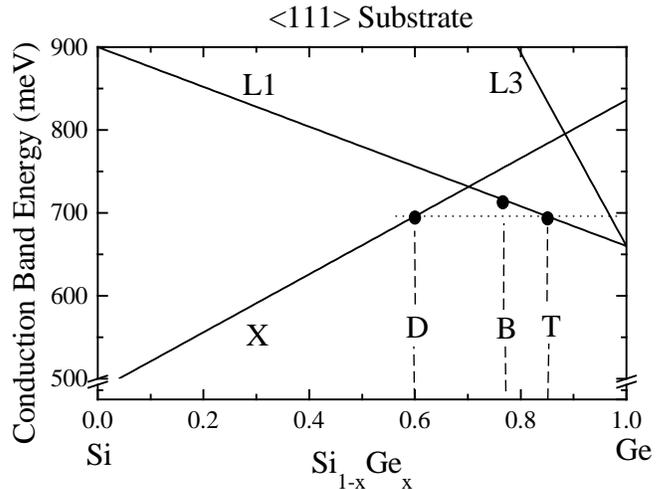}
}
\vspace{-35ex}
\caption{The conduction band energy in Si-Ge alloys, compositionally strained in 
the $\langle 111 \rangle$ direction, from neutral strain at 100\%~Ge.  The X-valley has 
6~minima that remain degenerate.  The L-valley has 4~minima that are split between 
L1 and L3.  The conduction band changes from the X- to L1-character at a composition 
of Si$_{0.3}$Ge$_{0.7}$. At this band transformation, the $xy$-effective mass becomes 
relatively light, the Bohr radius increases, and the $g$-factor drops from $g \approx 1.998 $ 
to $g=g_{\parallel} \approx 0.823$.  
The fractional compositions D, T, and B, will be used 
in our band structure engineered, spin resonance transistor.}
\label{fig:fig1}
\end{figure}
Under 
$\langle 111 \rangle$ strain the 4~conduction band valleys split so that one of 
them is lowest in energy and is labeled L1.  The other 3~valleys remain 
degenerate and are labeled L3.  
Figure~\ref{fig:fig1} shows the conduction band 
structure in the Si-Ge alloys, grown compositionally strained in the $\langle 
111 \rangle$ direction, with neutral strain at 100\%~Ge, as adapted from a more 
complete set of band structures from Wang {\it et al}\cite{wang-93}.

The hydrogenic Schr\"{o}dinger equation for anisotropic effective mass, $m_{xy}$~in 
the plane of the wafer, and $m_z$~perpendicular to the plane of the wafer, has 
been solved for arbitrary values of $m_{xy}/m_z$ by 
Schindlmayr\cite{schindlmayr-97}.  The Bohr radius in the $xy$-plane is 
influenced by both effective masses: 
\begin{equation}
\label{eq:bohrradius}
a_{B,xy}=\frac{2 \epsilon}{3 \pi} \frac{2+(m_{xy}/m_z)^{1/3}}{m_{xy}}a^0_B
\end{equation}
\begin{table}[h]    
\caption{Conduction band effective masses relative to m$_0$, and the corresponding 
Bohr radii and $g$-factors in Si and Ge.
\label{table:table1}
}
\begin{center}
\begin{tabular}{llllllll}  
material  & $\epsilon$ & $m_{xy}$ & $m_{z}$ & $a_{B,xy}$ & $a_{B,z}$ & $g_{\parallel}$ & $g_{\perp}$ \\[0.5ex] \hline
Germanium & 16         & 0.082    & 1.59    & 64 \AA	   & 24 \AA    & 0.823		  & 1.933	  \\
Silicon   & 12         & 0.191    & 0.916   & 25 \AA     & 15 \AA    & 1.999		  & 1.998     \\
\end{tabular}
\end{center}
\end{table}
with $a^0_B$ the Bohr radius of a free hydrogen atom and $m_{xy}<<m_z$ is 
assumed, as is appropriate for the z-oriented Si~and~Ge conduction band 
ellipsoids.  The Bohr radius in the heavy mass direction, $a_{B,z}$ is given by 
$a_{B,z}=(\frac{m_{xy}}{m_z})^{1/3}a_{B,xy}$. Using the actual masses and the 
exact formula\cite{schindlmayr-97}, we give the 
Bohr radii in Si~and~Ge for $z$-oriented conduction band ellipsoids 
in Table~\ref{table:table1}.  
\begin{figure}[h]
\centerline{\epsfxsize=250pt
		\epsfbox{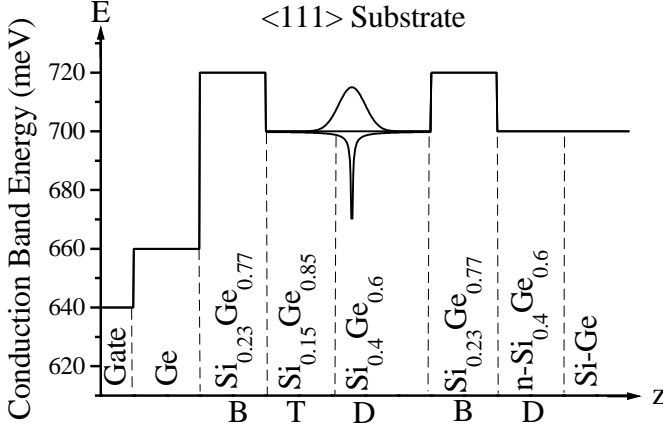}
}
\vspace{-37ex}
\caption{The band structure diagram for the proposed spin-resonance transistor, 
showing the Coulombic potential well of the donor ion in the Si$_{0.4}$Ge$_{0.6}$~D-layer where 
the conduction band minimum is X-like.  The hydrogenic wave function partly overlaps 
the Si$_{0.15}$Ge$_{0.85}$~T-layer where the conduction band minimum is L-like.  
The donor electron is confined by the two Si$_{0.23}$Ge$_{0.77}$~B-barrier layers.  
The epilayer thicknesses are not to scale.}
\label{fig:fig2}
\end{figure}

In Table~\ref{table:table1}, special note should be taken of the Bohr 
radius of 64~\AA~for $\langle 111 \rangle$~strained Ge-rich alloys in which 
the L1~band minimum forms the conduction band.  At that orientation, the X-band 
minima in Si-rich alloys would have a Bohr radius of only~$\approx 20$~\AA.  
Thus we achieve over a factor~3 increase in the transistor spacing by using a 
Ge-rich layer.

Given that the exchange interaction is a dominant influence among the donor 
spins, we make the point that Preskill's de-coherence criterion can be 
redefined\cite{gea-98} as the on/off ratio of the spin-spin interaction, as 
induced by the transistor gates.  The actual required transistor spacing is set 
by the need for the weakest possible exchange interaction when the 2-qubit 
interaction is off, and a strong exchange interaction when 2-qubit interactions 
are turned on.  The exchange energy $4J$ between hydrogenic wave functions 
determines both time scales:
\begin{equation}
\label{eq:exchange}
\frac{4 J(r)}{h} \approx 1.6 \frac{q^2}{\epsilon a_B}(\frac{r}{a_B})^{5/2}\exp{\frac{-2r}{a_B}}
\end{equation}

If we require the exchange energy in the off-state to be less than the 
measured\cite{chiba-72} T$_2$~dephasing linewidth $\approx$1~kHz, then the donor 
ions would have to be about 29~Bohr radii apart, allowing a spacing of about 
$2000$~\AA.  Such critical dimensions are well within the range that can be 
produced by electron beam lithography.  

Later we will show that by gate-controlled Stark distortion of the hydrogenic 
wave functions, the Bohr radius can be further increased, switching on the 2-
qubit interactions.  Thus, band structure engineering allows us to use only one electrostatic
gate to control both one- and two-qubit operations, rather than two separate A- and J-gates as 
required by Kane. This reduction of the number of gates by a factor of two, though not 
essential for the operation of the our ESR, means that all lithographic dimensions are doubled,
which significantly increases the manufacturability of the device.

\section{Gate Controlled Single Qubit Rotations in the Spin-Resonance 
Transistor}
\label{sec:onequbit}
The essence of a spin-resonance transistor (SRT) qubit is that a gate electrode 
should control the spin-resonance frequency.  By tuning this frequency with respect 
to the frequency of a constant radiation field, that is always present while the
computer is being operated, single qubit rotations can be readily implemented on 
the electron spin.  A band structure diagram for 
the SRT is shown in Figure~\ref{fig:fig2}.

\begin{figure}[h]
\centerline{\epsfxsize=250pt
		\epsfbox{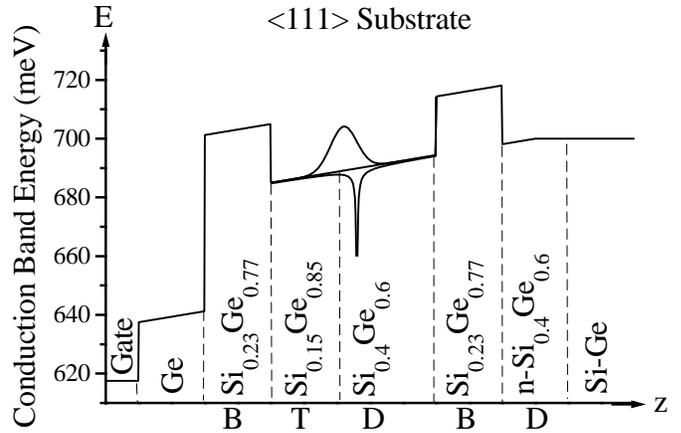}
}
\vspace{-37ex}
\caption{The donor electron wave function is electrostatically attracted toward the 
Si$_{0.15}$Ge$_{0.85}$~T-layer where the 
conduction band minimum is L1-like.  There it will experience a smaller $g$-factor, 
that is gate tunable.  The actual $g$-factor will be a weighted average between the D-
and T-layers.}
\label{fig:fig3}
\end{figure}
We rely on the difference in electronic $g$-factor, $g=1.998$ for Si-rich 
alloys, and $g=g_{\parallel}=0.823$ for Ge-rich alloys, strained in the $\langle 111 \rangle$ 
direction.  
Thus, the electron spin resonance 
transition can be readily tuned by an electrostatic gate on a compositionally 
modulated Si-Ge epilayer structure, such as shown in Figure~\ref{fig:fig3}.  
In a 
study of the composition dependence of the $g$-factor in Si-Ge alloys, Vollmer 
and Geist\cite{vollmer-74} showed that the $g$-factor is most influenced by the 
band structure crossover from X~to~L1 at a composition of Si$_{0.3}$Ge$_{0.7}$, 
and hardly at all by compositional changes away from that crossover.  
The $^{31}$P dopant atoms are positioned in the Si$_{0.4}$Ge$_{0.6}$~D-layer, 
a composition which is to the left of the crossover in Figure~\ref{fig:fig1}.
By electrostatically attracting the electron wave function into the 
Si$_{0.15}$Ge$_{0.85}$ T-layer, the spin resonance can be tuned very substantially. 
\begin{figure}[h]
\centerline{\epsfxsize=250pt
		\epsfbox{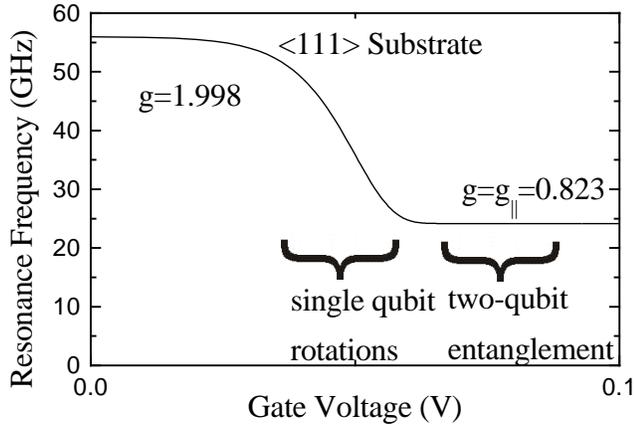}
}
\vspace{-35ex}
\caption{A schematic of the dependence of the spin resonance frequency on the 
transistor gate voltage. As the electrons are pulled toward 
the positive gate electrode and into the more Ge-rich alloy compositions, the 
hetero-barrier B-layer prevents the donors from 
becoming completely ionized.  At intermediate gate voltages, the $g$-factor can be 
tuned from $g$=1.998 to $g$=0.823.  The frequencies on the vertical axis correspond to 
a magnetic field of 2~Tesla.  The two-qubit tuning range will be explained in the next 
section.}
\label{fig:fig4a}
\end{figure}    

The two barrier layers of composition Si$_{0.23}$Ge$_{0.77}$, labeled B in 
Figure~\ref{fig:fig2}, have a conduction band structure as indicated in 
Figure~\ref{fig:fig1}.   They have an L1-like conduction band minimum, to the 
right of X-L1 band structure cross-over, and thus have the same $g$-factor as 
the Si$_{0.15}$Ge$_{0.85}$ T layer. The purpose of the B~layers is to confine 
the donor electrons and prevent them from tunneling away and becoming lost.  The 
energy height of the barrier need only be comparable to the donor binding 
energy, $\approx 20$meV to fulfil this task.  On the other hand the 
Si$_{0.4}$Ge$_{0.6}$ D-layer and the Si$_{0.15}$Ge$_{0.85}$ T-layer should have 
no energy barrier between them so that the $g$-factor can be freely tuned.  
Thus 
the D layer and the T layer are selected at compositions straddling the X-L1 
crossover in Figure~\ref{fig:fig1}, so that their respective conduction band 
energies $E_D$ and $E_T$ are the same.  A schematic tuning curve for our 
proposed spin resonance transistor is shown in Figure~\ref{fig:fig4a}.
As the spin resonance transistors are tuned in and out of resonance with the 
radiofrequency field the electron spin can be flipped, or subjected to a phase change.

\begin{figure}[h]
\centerline{\epsfxsize=250pt
		\epsfbox{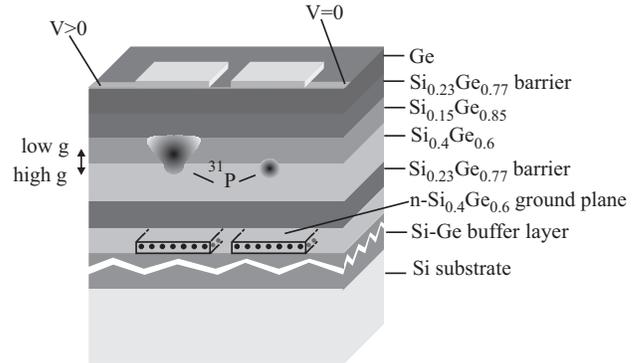}
}
\vspace{-47ex}
\caption{The left transistor gate is biased $V > 0$ producing single qubit unitary 
transformations in the left SRT.  The right gate is unbiased, $V=0$.  The 
n-Si$_{0.4}$Ge$_{0.6}$ ground plane is  counter-electrode to the gate, and it also 
acts as an FET channel for sensing the spin.}
\label{fig:fig7}
\end{figure}

The wave function distortion during tuning is shown for the 
left side transistor in Figure~\ref{fig:fig7}.
The confinement barriers of composition~B Si$_{0.23}$Ge$_{0.77}$, play an 
important role.  They must confine the qubit donor electrons for long periods of 
time, or the carriers and their quantum information will be lost.  For that purpose 
the B-barrier layers each need to be about $200$~\AA~thick, for a carrier 
lifetime comparable to the $\approx$1hour T$_1$~spin-lattice relaxation for electron spin flips.
The two layers combined would total about $400$~\AA, 
well within the practical strain limit\cite{wang-95} of $\approx 1000$~\AA~for 
growth of a~23\%~compositionally strained alloy. The D~and~T layers have 
thicknesses similar to the $a_{B,z}$ vertical Bohr radius and contribute only 
slightly to the strain burden.  

\begin{figure}[h]
\centerline{\epsfxsize=250pt
		\epsfbox{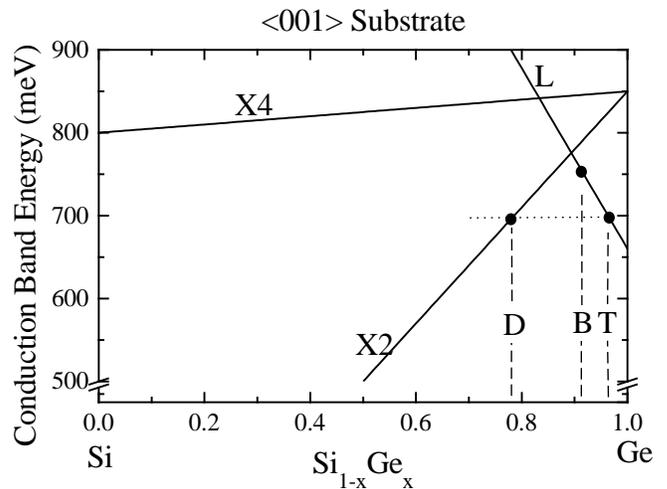}
}
\vspace{-35ex}
\caption{The conduction band energy in Si-Ge alloys, compositionally strained in the 
$\langle 001 \rangle$ direction, from neutral strain at 100\%~Ge.  The L-valley has 4 
minima that remain degenerate. The X-valley has 6 minima along the cubic directions,
that are split between X4 and X2. The compositions 
D, T, and B are much less strained than in the $\langle 111 \rangle$ case, and allow 
for higher barrier heights to confine the dopant electron. For this crystal orientation, the 
$g$-factor in the Ge-rich T- and B-layers is $g=1.563$ }
\label{fig:fig5}
\end{figure}  

\begin{figure}[h]
\centerline{\epsfysize=92mm
		\epsfbox{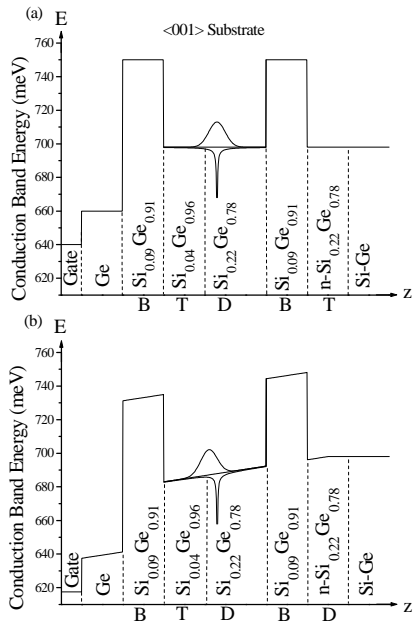}
}
\caption{The band structure diagram for the spin-resonance transistor, with epilayers 
grown in the $\langle 001 \rangle$ direction.  Both the unbiased (a) and the biased 
case (b) are shown. The conduction band energies allow the selection of layers with 
composition D, T, and B such that the confining barrier height is increased to 50~meV, 
while the strain in the layers is reduced, compared to the $\langle 111 \rangle$ 
orientation.  The epi-layer thicknesses are not to scale.}
\label{fig:fig6}
\end{figure}

If one uses alloys grown in the $\langle 001 
\rangle$ direction instead, the numbers become even more favorable.  
Figure~\ref{fig:fig5} shows the conduction band structure in the Si-Ge alloys, 
grown in the $\langle 001 \rangle$ direction\cite{wang-95}, compositionally 
strained from neutral strain at 100\%~Ge.  In this growth direction, the L~band 
remains unsplit, and the X~band splits up into a doubly degenerate X2 and a 
quadruply degenerate X4 band.  As can be seen, the conduction band energy 
changes much more rapidly as a function of alloy composition for the 
$\langle 001 \rangle$ growth 
direction.  Moreover, the X2 and the L bands cross over at approximately 90\%~Ge 
instead of 70\%~as in the Ge $\langle 111 \rangle$ case.  This allows us to select 
alloys with much lower strain, 
while obtaining a barrier height of 50~meV, more than twice the barrier height 
obtained in the $\langle 111 \rangle$ direction. 
Consequently the layers can be 
made thinner while still preventing tunneling of the dopant electron and the 
strain tolerance is significantly improved. The corresponding band 
structure diagram for the $\langle 001 \rangle$ oriented SRT is shown in Figure~\ref{fig:fig6}. 

\begin{figure}[h]
\centerline{\epsfxsize=250pt
		\epsfbox{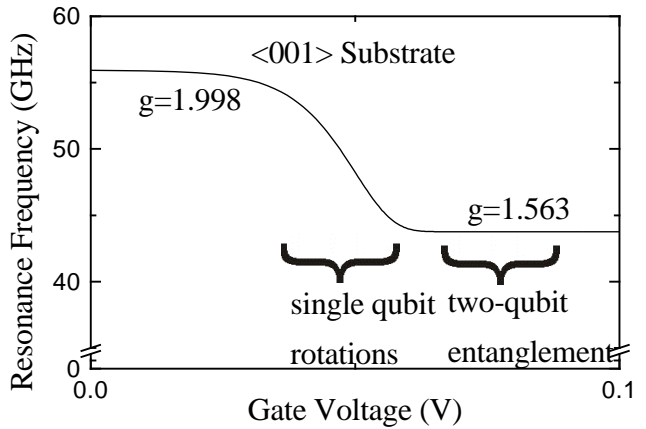}
}
\vspace{-35ex}
\caption{A schematic of the dependence of the spin resonance frequency on the 
transistor gate voltage for the case of a $\langle 001 \rangle$ substrate. The static magnetic
field is in the $\langle 001 \rangle$ direction and has a strength of 2 Tesla. 
The tuning range is reduced in this growth direction with respect to the $\langle 111 \rangle$
case, because the $g$-factor in the Ge-rich layer is different: $g=1.563$.}
\label{fig:fig4b}
\end{figure}

\noindent
However, in the $\langle 100 \rangle$ direction, the $g$-factor is equal to the average
value: $g=1.563$, so that the tuning range for the spin resonance frequency is less than in 
the $\langle 111 \rangle$ case, as is demonstrated in Figure~\ref{fig:fig4b}.

The use of the $\langle 001 \rangle$ growth direction comes at the expense of an 
increased effective mass in the $xy$-plane and a lighter mass in the $z$-direction. 
The conduction band ellipsoid pointing in the $\langle 111 \rangle$ direction is 
55\degr away from the $\langle 001 \rangle$ direction and thus the $z$-direction 
no longer coincides with the heavy mass direction ($\langle 111 \rangle$). 
Some of the heavy mass is transferred into the $xy$-plane, resulting in shorter 
Bohr radii. However, the lightest mass in Ge is equal to the heaviest mass in 
Si(see Table~\ref{table:table1}). Therefore, 
the Ge-rich layer will always remain the layer with Bohr radii in the $xy$-direction 
which are at least as large as those in the Si-rich layer. Therefore Ge-rich layers 
will again perform the function of the tuning T-layer, and the barrier B-layer for 
structures grown in the $\langle 001 \rangle$ as they did for the $\langle 111 \rangle$ 
direction.

\section{Two-qubit interactions}
\label{sec:twoqubit}
The spin resonance transistors must be spaced far enough apart, that they will 
not produce phase errors in one another.  At the same time it is necessary to 
allow wave function overlap for the exchange interaction to activate the 2-qubit 
interactions.  These are needed to produce for example a Controlled NOT (CNOT) 
gate, which is required to build a universal set of quantum logic gates.  To 
achieve this we rely on our ability to tune the Bohr radius of the donors in the 
$xy$-direction parallel to the semiconductor surface.  

The Bohr radius $a_B$ of a hydrogen-like donor increases with decreasing 
binding energy. A famous example is excitons confined in a 2-d flat quantum well:  
The excitonic binding energy is four times greater\cite{bastard-81} than it 
would be in 3 dimensions.  The reason is that spatial confinement forces the 
electron to spend more time near the positive charge, and it experiences tighter 
binding.  Accordingly the Bohr radius is diminished.  For the same reason, 
confinement by heavy mass in the z-direction reduces the Bohr radius in the 
$xy$-plane as can be seen from Equation~\ref{eq:coulomb}.  Without this reduction the effective mass 
in the $xy$-direction in strained $\langle 111 \rangle$ Ge would even be higher. 

Our technique for 2-qubit interactions does not require any J-gates. By increasing the
gate voltage, we pull the electron wave function 
away from the positive ion, to reduce the binding energy, and increase the wave 
function overlap between electrons bound to neighboring dopant ions.  As shown in 
Figure~\ref{fig:fig3}, the electrons can be 
electrostatically attracted to one of the barriers formed by the 
Si$_{0.23}$Ge$_{0.77}$~B-composition layer, forming a type of modulation doped 
channel in the $xy$~plane.  The binding energy to the positive ions is greatly 
weakened, since the electrons are spending most of their time near the 
Si$_{0.23}$Ge$_{0.77}$~B-barrier.  
\begin{figure}[h]
\centerline{\epsfxsize=250pt
		\epsfbox{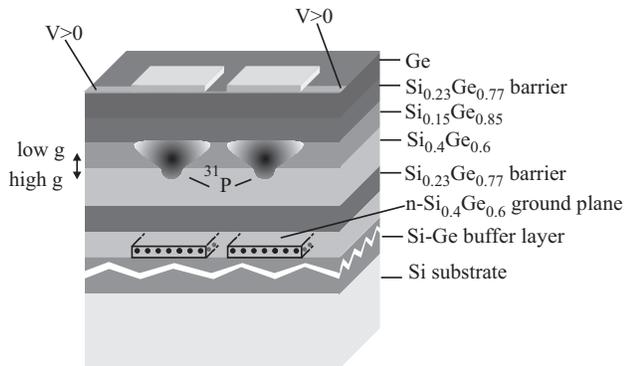}
}
\vspace{-47ex}
\caption{Attracting the electrons to the 
Si$_{0.23}$Ge$_{0.77}$ B-barrier reduces 
their Coulomb binding energy and increases their wave function overlap, allowing 
2-qubit interaction.}
\label{fig:fig8}
\end{figure}
\noindent
Consequently the Coulomb potential becomes weakened to the following form:
\begin{equation}
\label{eq:coulomb}
V=-\frac{1}{4 \pi \epsilon_0 \epsilon} \frac{q}{\sqrt{r^2+d^2}}
\end{equation}
where $r^2=x^2+y^2$ is the horizontal distance from the donor ion, squared, and 
$d$ is the vertical spacing from the barrier to the donor ion, and $q$ is the 
electronic charge.
Thus by adjusting the vertical depth of the ion, $d$, the 
Coulomb potential can be made as weak as desired.  
The weak Coulomb binding 
energy implies a large Bohr radius. The large radius permits a substantial wave 
function overlap in the $xy$-plane along the B-barrier layer, and a substantial 
2-qubit exchange interaction.  It should be possible to tune from negligible 
exchange interaction, all the way to a conducting metallic 2-d electron gas, by 
adjusting the vertical spacing $d$.  As the electrons overlap, they 
will interact through the exchange interaction. It was already shown by 
DiVincenzo\cite{loss-98}, that the exchange interaction can produce CNOT quantum 
gates.  

The gate bias voltage range for 2-qubit entanglement, is indicated by the second 
curly bracket in Figure~\ref{fig:fig4a}.  That voltage range attracts the 
electrons away from the positive ions and toward the Si$_{0.23}$Ge$_{0.77}$~B 
barrier, thus increasing their wave function overlap.  In the mid-voltage range, 
the first curly bracket in Figure~\ref{fig:fig4a}, 1-qubit rotations take 
place.  Thus both one- and two-qubit interactions can be controlled by a single 
gate.  Gate tuning of 
a 2-qubit exchange interaction is illustrated in Figure~\ref{fig:fig8}. 

\section{Detection of Spin Resonance by a FET Transistor}
\label{sec:detection}
\begin{figure}[h]
\centerline{\epsfxsize=250pt
		\epsfbox{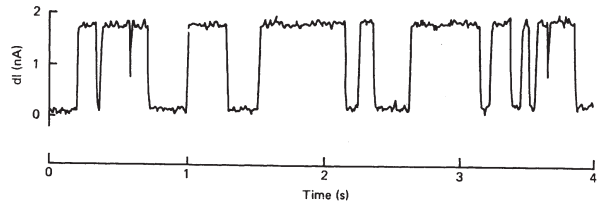}
}
\vspace{-62ex}
\caption{
The current noise in a small FET at 83~K from Kurten {\it et al}\protect\cite{kurten-89}. 
At this temperature the channel current fluctuates between two states, caused by a 
single trap being filled and emptied by a single charge.  The change in channel current 
is $\approx$2nAmps, which represents a few percent of total channel current, and is 
easily measured.
}
\label{fig:fig9}
\end{figure}
It is a truism of semiconductor electronics that we need crystals of high 
perfection and extraordinary purity.  Semiconductor devices are very sensitive 
to the presence of chemical and crystallographic faults down to the level of 
10$^{11}$~defects/cm3 in the volume, and 10$^8$~defects/cm2 on the surface.  
Such defect concentrations are far below the level of sensitivity of even the 
most advanced chemical analytical instruments.  These imperfections influence 
the electrical characteristics of semiconductor devices, as they vary their 
charge states.  Thus conventional electronic devices are sensitive to very low 
concentrations of defects.  

The detection sensitivity becomes particularly striking when the electronic 
devices are very tiny, as they are today.  If electronic devices are small 
enough, then there is a good probability that not even one single defect might 
be present in, or on, the device.  That helps define the potential yield of 
essentially perfect devices.  But if a defect were to be present, it would have 
an immediate effect on the current-voltage (I-V) characteristics of that device.  
Therefore, the new world of small transistors is making it relatively easy to 
detect single defects, as their charge states directly influence the I-V curves.  

As Kane pointed out, the essential point for us is to detect spin, not by its miniscule 
magnetic moment, but by virtue\cite{kane-98} of the Pauli Exclusion Principle.  
A donor defect can bind\cite{taniguchi-77} a second electron by 1meV, provided that 
second electron has opposite spin to the first electron.  
Thus spin detection 
becomes electric charge detection, the essential idea\cite{loss-98} behind Spin Resonance 
Transistors.  In a small transistor, even a single charge can be relatively 
easily monitored.  

A fairly conventional, small, Field Effect transistor, (FET) is very capable of 
measuring single charges, and therefore single spins as well.  A single 
electronic charge, in the gate insulator, can have a profound effect on a low 
temperature FET.  At more elevated temperatures for example, the motion of such 
individual charges produces telegraph noise in the FET channel current.  An 
illustration of such single charge detection\cite{kurten-89} is in 
Figure~\ref{fig:fig9}. A single electrostatic charge can add 1~additional carrier to the few hundred 
electrons in a FET channel.  However the 2 nAmp change in channel current seen 
in Figure~\ref{fig:fig9} represents a few percent change, and is caused by long 
range Coulomb scattering influencing the resistance seen by all the electrons.  
At low FET operating temperatures, $\approx$1~K, the random flip-flops disappear, 
but the sensitivity to single charges remains\cite{comment-Eli}. 

In our spin-resonance transistor design, shown in Figures~\ref{fig:fig7} and 
\ref{fig:fig8}, the FET channel is labeled as the n-Si$_{0.4}$Ge$_{0.6}$~ground 
plane counter-electrode.  It is located under the $^{31}$P qubit donor, and in 
turn, the donor is under the top surface gate electrode.  Thus the spin qubit is 
sandwiched between two electrodes.  As in a normal FET the gate electrode 
modulates the n-Si$_{0.4}$Ge$_{0.6}$~channel current.  The qubit electron donor 
is positioned in the gate insulator region where its charge state can have a 
strong influence on the channel current.  Thus the successive charge states: 
ionized donor, neutral donor, and doubly occupied donor (D$^-$ state) are 
readily sensed by measuring the channel current.

\begin{figure}[h]
\centerline{\epsfxsize=250pt
		\epsfbox{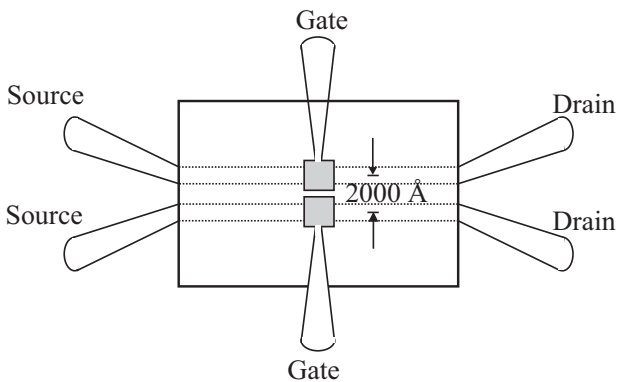}
}
\vspace{-45ex}
\caption{Top view of the proposed device to demonstrate a CNOT gate.
A perspective view (, not including the source and drain,) is shown in 
Figure~\ref{fig:fig7}. Fluctuations in the current that flows from source to
drain signal the charge state of the dopant ion under each electrode.
}
\label{fig:fig10}
\end{figure}

In Figures~\ref{fig:fig7} and \ref{fig:fig8}, the two transistors have separate 
sensing channels under each transistor, so that they can be separately 
monitored, or indeed monitored differentially.  By adjusting the gate 
electrodes, both qubit donor electrons can be attracted to the same donor.  If 
they are in the singlet state they can join together forming the D- state on one 
of the two dopant ions, but in the triplet state they could never occupy the 
same site.  

Since the D$^-$ state forms on one transistor, and an ionized donor D$^+$, on 
the other transistor, there would be a substantial change in differential 
channel current to identify the singlet state.  For the triplet state, both 
donors remain neutral and differential channel current would be constant.  As 
indicated by the caption to Figure~\ref{fig:fig9}, we can anticipate a few 
percent change in FET current associated with the singlet spin state, making spin 
readily detectable.  

\section{Small scale demonstration}
\label{sec:smallscale}
A possible 2-qubit demonstration device is shown in Figure~\ref{fig:fig10}. The 
differential current between the two FET's channels in 
Figure~\ref{fig:fig8} would monitor the electron spin resonance. In 
practice a large number of transistor pairs would be arrayed along the two FET 
channels in Figure~\ref{fig:fig10}, to allow for a finite yield in getting 
successful pairs.  A good pair can be sensed using the same technique used in the previous
section for the detection (measurement) process.

There are two levels of doping in our proposed device: The first level of doping 
is the conducting FET channel doping, that needs to be at a heavy concentration to overcome 
freeze-out at low temperatures.  This is a standard design technique in low 
temperature electronics. The second level of doping is in the qubit layer, that 
allows only one donor ion per transistor. Both doped regions need to be 
spatially patterned.  The doped layers can be implemented by conventional ion-
implantation through a patterned mask, possibly with an intermediate epitaxial 
growth step to minimize ion straggle.  Conventional annealing can be used to 
remove ion damage.

\begin{figure}[h]
\centerline{\epsfxsize=250pt
		\epsfbox{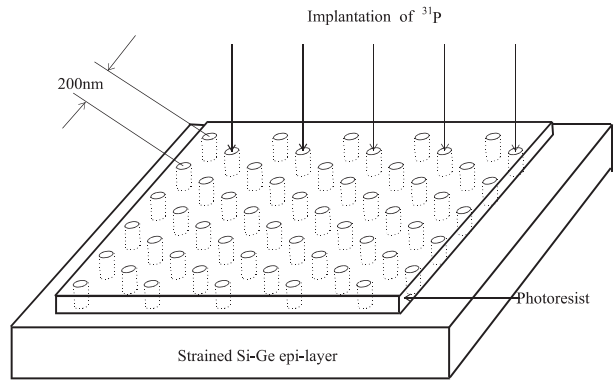}
}
\vspace{-43ex}
\caption{The ion implantation step for inserting an array of qubit donor ions.  
The buried FET channels, that act as counter-electrodes to the gates and sense the 
spin/charge state, would be produced the same way.  In a small-scale demonstration, 
the array would consist of only 2~rows, aligned with the FET channels of 
Figure~\ref{fig:fig10}. This should provide an adequate yield of good qubit pairs.}
\label{fig:fig11}
\end{figure}
The ion-implantation dose for the qubit layer would be adjusted so that on 
average, only 1~Phosphorus ion would fall into each opening in the photoresist 
layer of Figure~\ref{fig:fig11}.  By Poissonian statistics, the probability of 
getting exactly 1~Phosphorus ion is 36.7\%.  Thus the probability of getting two 
adjacent gates to work would be 13.5\%.  That is adequate yield for a small-scale
two-qubit demonstration device.  To improve the yield for scale-up, there are many 
options.  For example, the dopant could be sensed by its electric charge, and 
re-implanted if it were absent.  Sensing an individual dopant is not difficult. 
It can be done, for instance,
by monitoring the I-V curve at each site.
By changing the voltage on a particular A gate the electrons can be
stripped off the donor. As result one can see no-change, a single-change,
or a double-change of the current depending on whether there is no donor,
one donor or two donors (etc.) in that site.

\section{Scaling Up}
\label{sec:scaleup}
There are a number of potential problems in scaling to a large computer.  The 
future usefulness of electron spins will depend heavily on the favorable 
homogeneous T$_2$~spin echo linewidth\cite{chiba-72} in Silicon, only 10$^3$Hz.  
The T$_2$~lifetime in Si-Ge alloys has not been measured, and it will have to be 
demonstrated that it is as favorable as in pure Silicon.  On the other hand 
there also appear to be methods such as isotopic purification, whereby this 
linewidth can be improved, 
particularly for well-isolated electrons.

\begin{figure}[h]
\centerline{\epsfxsize=250pt
		\epsfbox{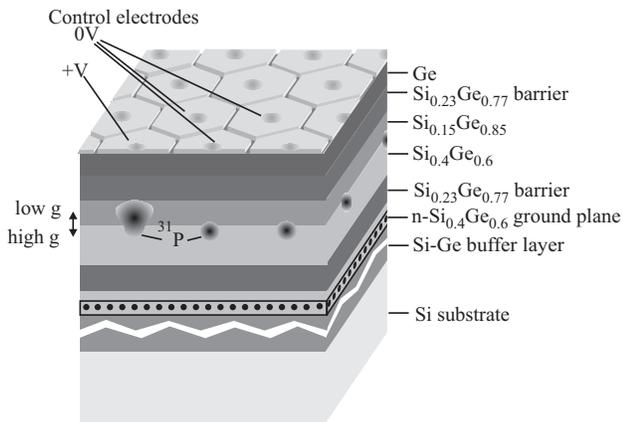}
}
\vspace{-40ex}
\caption{In the future, we can expect arrays of Si-Ge SRT transistors.  The 
center-to-center spacing would be $\approx 2000$~\AA.  
The gate electrodes on top will perform both single and 2-qubit operations, and can be 
used for data and instruction read-in.}
\label{fig:fig12}
\end{figure}

In very large arrays, there are problems associated with the implantation yield of 
qubit donors.  Poissonian statistics gives a yield of 36.7\%, while a yield of 
50\%~will required for percolation, or quantum connectivity, through the two-
dimensional triangular array.  There have been numerous non-Poissonian doping schemes 
proposed including sense/re-implant, self-assembly of molecular dopants, and 
scanning probe writing.  Innovative doping methods have a long history, and we 
should anticipate that a suitable method will be optimized in time for scale-up 
to large quantum computers.

For instance, the sense/re-implant method (in which empty sites are
sensed, and re-implanted with doping probability $p_n$ in the $n$'th implant)
yields $p e^{-p} ( 1 - e^{-np})/(1 - e^{-p})$ good sites when $p_i = p$ is
chosen. With this formula, already $n=2$ (only one additional implant)
passes the percolation limit to yield
$52.16\%$, while more implants, n=3,5,9, and 24,
yield more than $60\%$, $70\%$, $80\%$, and $90\%$
good sites respectively.
With $n=2$ an optimization of the doping probability in each implant
(to be $p_1 = 0.632$ and $p_2 = 1$) provides the 
optimal yield of $53.15\%$.

The other scale up issue revolves around the fact that each transistor will not 
be identical.  As Kane noted, the transistors will have to be checked and calibrated 
repeatedly for use in a full-fledged quantum computer.  The reason is that the nuclear 
spins, although almost static, will be different for each transistor.  In addition 
the local alloy structure is different near every donor.  We should not be 
discouraged by this checking and calibration requirement.  In manufacturing 
classical integrated circuits, testing and repair are the biggest expense.  It is common to 
have only a finite yield of good devices, and to reroute wiring around bad 
transistors.  This is probably inherent in the manufacture of any large-scale 
system.

The size of spin resonance transistors, the required defect density, the 
increasing use of Si-Ge alloys, are all near to the present state of technology.  
If the spin resonance transistor (SRT) is successfully developed, we can 
anticipate arrays of qubits appearing much as in Figure~\ref{fig:fig12}.

\begin{figure}[h]
\centerline{\epsfxsize=250pt
		\epsfbox{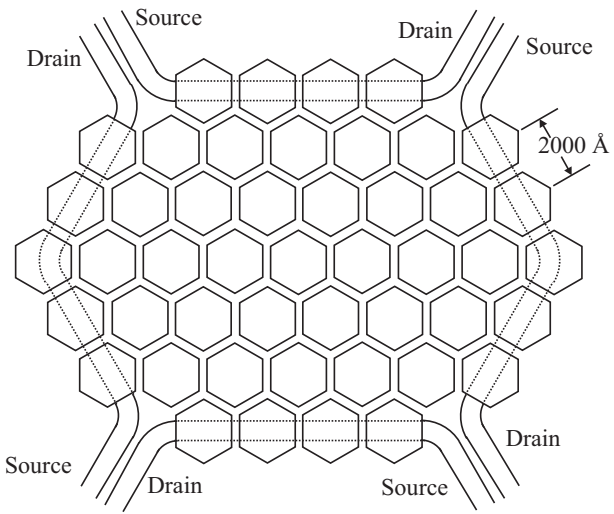}
}
\vspace{-35ex}
\caption{In a large array, the read-out qubits would be located around the periphery.  
Buried FET channels would sense the spin/charge state of a selected qubit.  
The channel current can change by a few percent in response to a single electronic 
charge.}
\label{fig:fig13}
\end{figure}
The read-out of data requires that the buried counter electrode, opposite the 
gate, should also function as an FET channel.  In a quantum computer, the result 
of the quantum computation is usually displayed on a small sub-array of all the 
qubits.  Hence the read-out qubits can be located at the edge of the array. 
Figure~\ref{fig:fig13} shows a qubit array, with read-out FET channels (counter-
electrodes) buried under the peripheral qubits of the array. A single buried FET 
read-out channel can serve many qubits, since a chosen qubit can be selected for 
readout by its gate electrode.  

\begin{figure}[h]
\centerline{\epsfxsize=250pt
		\epsfbox{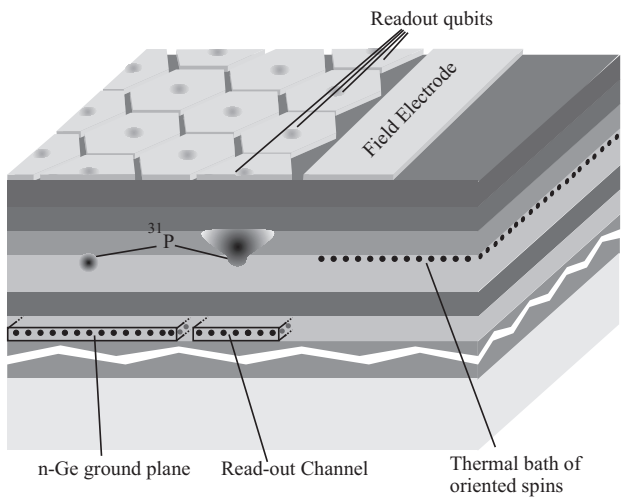}
}
\vspace{-33ex}
\caption{A perspective view of Figure~\ref{fig:fig13}, gives more details of the 
readout architecture for the peripheral qubits.  The field electrode allows the 
Readout Qubits to interact with the heat bath of oriented electron spins}
\label{fig:fig14}
\end{figure}

The read-out operation can be expedited if there is a thermal reservoir of 
donors surrounding the peripheral qubits as shown in Figure~\ref{fig:fig14}.  
These can be attracted by a field electrode to the Si${_0.23}$Ge$_{0.77}$~B-barrier 
under the electrode, forming in effect a modulation doped layer.  Since 
the operating temperature of the computer is such that $kT \ll E_z$ with $E_z$ the 
Zeeman energy of the electron spins, these qubits would be oriented by the 
magnetic field, and would act as a spin heat bath of known orientation.  By 
attracting those bath spins to a peripheral read-out qubit gate electrode, a singlet 
state could be formed, sensing that the readout qubit had been flipped.  The 
current in the FET channel would then change, completing the read-out operation.  
\noindent

After readout, the gate voltage could be made even more positive, and the read-
out qubit could thermalize with the surrounding heat bath.  In effect, this 
resets the initial state of that peripheral qubit, which could then be swapped 
into the interior qubits for re-use as fault correcting ancilla qubits. 

Without a doubt there will be many other issues regarding scale-up. 
Semiconductors, particularly silicon, provide a track record of being tractable, 
engineerable materials in which many difficult accomplishments have become 
routine.


\end{document}